# Low Complexity Fair Scheduling in LTE/LTE-A Uplink Involving Multiple Traffic Classes

Atri Mukhopadhyay, Goutam Das

*Abstract*— **The bulk of the research on Long Term Evolution/Long Term Evolution-Advanced packet scheduling is concentrated in the downlink and the uplink is comparatively less explored. In uplink, channel aware scheduling with throughput maximization has been widely studied while considering an infinitely backlogged buffer model, which makes the investigations unrealistic. Therefore, we propose an optimal uplink packet scheduling procedure with realistic traffic sources. Firstly, we advocate a joint channel and buffer aware algorithm, which maximizes the actual transmitted bit-count. Thereafter, we introduce delay constraints in our algorithm to support real-time traffic. We further enhance our algorithm by incorporating the varied delay and throughput requirements demanded by mixed traffic classes. Finally, we introduce priority flipping to minimize bandwidth starvation of lower priority traffic in presence of higher percentage of high priority traffic. We observe that a delay constraint may render the optimization-based proposals infeasible. Therefore, to avoid infeasibility, we replace the delay constraint with delay outage minimization (DOM). DOM aims at minimizing the packet drop due to delay violation. Moreover, DOM also helps in reducing the problems to a well-known assignment problem, which can be solved by applying the Hungarian algorithm. Hence, our approach delivers an optimal allocation with low computational complexity.**

*Index Terms*— **Assignment, fairness, knapsack, LTE, LTE-A, MAC throughput, priority flipping, uplink scheduling**

## I. Introduction

Mobile communication has progressed in leaps and bounds over the past few years fueled by the high data rate demand from the mobile users. Long term evolution (LTE) by the third generation partnership project (3GPP) [1] and LTE-Advanced (LTE-A) [2][3], also known as pre-fifth-generation (5G) are the most promising mobile technologies that provide per user data rates of the order of megabits per second [1]. LTE-A introduces carrier aggregation [2][3] and guarantees seamless integration of voice, video and data. The LTE/LTE-A physical layer is fundamentally different and uses orthogonal frequency division multiple access (OFDMA) in its downlink and single carrier frequency division multiple access (SC-FDMA) in its uplink [1]. Both OFDMA and SC-FDMA divide the channel into multiple sub-carriers. Allocating the sub-carriers to different users having different channel conditions to maximize system utilization a major challenge in LTE and LTE-A.

A. Mukhopadhyay and G. Das are with G. S. Sanyal School of Telecommunications, Indian Institute of Technology, Kharagpur, India. E-mail: atri.mukherji11@gmail.com.

The responsibility of packet scheduling (allocating a certain group of sub-carriers to a user) lies with the evolved nodeB (eNodeB). This unit of allocation is known as the physical resource block (PRB) that consists of a time duration of 1 ms defined as transit time interval (TTI) [4]. The eNodeB allocates multiple PRBs to multiple users with the target of maximizing the possibility of bit transmission in each and every TTI. Scheduling is done on both the downlink and the uplink. While downlink allows distributed allocation of PRBs [1][4]; the uplink of LTE requires a contiguous allocation of PRBs due to the restrictions imposed by SC-FDMA control overhead [4]. SC-FDMA also imposes a limit on the maximum number of users that can be scheduled at a time and commands the user to use a single modulation and coding scheme (MCS) in allocated PRBs within a certain TTI [5]. Though, downlink scheduling has been widely investigated, uplink scheduling is still comparatively less explored in both LTE and LTE-A.

The earliest approaches of LTE scheduling have targeted throughput maximization. Unfortunately, the algorithms existing in the literature consider infinitely backlogged model for data traffic and carry out only channel aware scheduling [6]-[13]. However, the MAC throughput, which is the actual number of bits transmitted, may remain low if actual buffer status is not considered along with the channel conditions while scheduling the user equipments (UEs) [14][15]. Hence, in this work, we have adopted a buffer and channel aware cross layer approach for optimal uplink scheduling.

Further, when real-time packets are scheduled, only MAC throughput maximization is not enough as the traffic these days mostly come with quality of service (QoS) constraints. The primary QoS constraints that need to be considered while scheduling are delay deadlines and packet drop constraints. In other words, the packets must reach the receiver within certain delay limits and the overall packet dropped must be within a specified threshold. Moreover, overall system QoS is lowered when fairness among the users are not looked into. However, throughput maximization on one hand and QoS and fairness consideration on the other hand are conflicting requirements and therefore, one has to settle for a tradeoff [16]. In order to get around the problem of meeting delay deadlines, proposals with a two-stage scheduling are available in the literature [12][17]. Here, in the first stage, time-domain packet scheduling (TDPS), the users are shortlisted based on head of the line (HoL) packet delay.

Thereafter, in the second stage, the frequency-domain packet scheduling (FDPS), the PRBs are allocated to the shortlisted users [12] to maximize the system throughput. Unfortunately, this leads to sub-optimal solutions [12].

Interestingly, optimal joint time and frequency allocation can be carried out by either enforcing a delay constraint towards achieving both throughput maximization and QoS improvement [18] or optimizing a multi-objective function [5]. Unfortunately, with a delay constraint, the optimization problem may become infeasible when the number of users that must be allocated to satisfy the delay constraint exceeds the number of available resource units. Multi-objective optimization, on the other hand, is computationally expensive [19][20] and requires proper parameter setting for providing weights to different objectives, which may differ for different network scenarios.

Apart from delay constraints, LTE uplink also needs to ensure sub-carrier contiguity constraint [5]. In the literature, there exist two distinct approaches to address this. The most popular approach is to follow PRB based allocation [5][21]-[24]. The PRB based scheduling approach requires optimization problem to enforce the PRB contiguity constraint. However, the contiguous PRB allocation restriction makes the assignment problem NP-Hard [6]. As a result, the proposals mostly come up with heuristic algorithms. Moreover, in such an allocation, the transmission power must also be adjusted depending on the number of PRBs allocated. Hence, the resulting optimization becomes computationally complex. Hence, the most effective approach is to group a set of adjacent PRBs and term them as resource chunks (RC) [6][7]. This approach carries out the allocation in terms of RCs and the MCS of the RC is equal to the minimum supported MCS by all the PRBs forming the RC. Even though this approach is heuristic, the advantages are that the scheduler no longer has to worry about the contiguous resource allocation. Also, the power adjustment is independent of allocation as the number of PRBs are fixed in an RC. Further, our previous results confirm that the RC based approach provides better results as compared to the other heuristic methods [14]. We, therefore, adopt the RC based approach in our framework.

Recently, in one of the proposed solutions to the uplink scheduling problem in LTE/LTE-A, named as adaptive and potential aware scheduling scheme (APASS), the authors have considered buffer and QoS aware throughput maximization [5]. APASS indirectly uses the RC approach and intend to keep flexible RCs. The proposal describes a three-stage procedure that carries out the PRB allocation for the traffic in a buffer while maintaining QoS. The final stage of APASS improves the scheduling performance of the initial allocation by following a potential zone concept. However, the user elimination method described in the second stage of APASS is heuristic and therefore, may result in a suboptimal allocation. Further, APASS has a comparatively high worst-case complexity[1] that might prohibit it to be used for a real time scheduler when the number of users is comparatively high. The paper enforces fairness by maintaining packet delay outage ratio (PDOR), which is the number of packets dropped due to delay overshoot divided by the total number of packets produced. However, this "reactive" approach considers the scheduling history and may not produce very promising results if the buffer condition is also considered. Ideally, one should proactively minimize the number of packets that will be dropped due to delay overshoot, if the user is not scheduled.

After carefully studying the literature, we conclude that a low complexity buffer aware optimal LTE uplink scheduler that simultaneously maximizes throughput and user fairness while reducing delay outage probability is still eluding. Hence, in this paper, we propose to design a novel scheduler that achieves the following objectives.

We modify the objective function of our optimization problem to include throughput achieved as well as throughput loss due to delay violation of packets if not scheduled. Therefore, the optimization simultaneously ensures that the throughput is maximized while number of bits dropped due to delay violation is also minimized. In the process, the delay deadline constraint for real-time traffic has been replaced by the number of packets dropped due to overshooting delay threshold in the objective function. We call this approach as delay outage minimization (DOM). Thereafter, we try to minimize the number of packets dropped due to delay threshold overshoot. This way, the problem is optimally reduced to a single objective optimization problem and at the same time, a possible infeasibility due to delay violation is avoided.

We further argue that our new objective function balances trade-off between the QoS and fairness issues. To be more elaborate, our approach converts packets to be dropped due to delay overshoot into a notion of lost throughput. Therefore, any user suffering from high packet loss will inadvertently reduce the throughput of the system. As a result, the suffering user will automatically be taken care by the system so that the overall throughput of the system is not hampered. Thus, our proposal also comes up with this unique way of ensuring fairness.

Finally, our optimization framework guarantees DOM in a proactive manner as it selects a user after looking into all the packets facing the danger of dropping and minimize this "packets to be dropped" metric. Thus, the proposed framework tries best to schedule packets even before they are dropped unlike existing literature where the scheduling parameters are tuned optimally in a reactive manner after the event of packet drop has happened [5].

At the end, in this paper, we have optimally transformed the formulated optimization problems to the well-known assignment problem of combinatorial optimization. Thereafter, we have solved them using the Hungarian algorithm [25], which gives an optimal solution. Our extensive simulation results reflect that the performances of our proposed algorithms are better than the state of the art.

In packet scheduling, the final frontier is reached when we try to deal with traffic of multiple classes. Different classes of real-time traffic have different delay requirements. In order to make our algorithm suitable for scheduling multiple classes of traffic, we have modified our input matrix of the optimization so that both delay sensitive traffic and best effort traffic can be handled simultaneously. The existing handling of multi class traffic often suffers from bandwidth starvation problem for the lower priority traffic. To alleviate this, we introduce the concept of priority flipping and show through extensive simulation that it can significantly improve the user experience.

[1] APASS has a worst-case complexity of $O(Umax(U,N)^3 log_2 N + NV)$, where $N$ is the number of resource blocks, $U$ is the number of active users and $V$ is the number of schedulable users.

The paper is organized as follows. Section II provides some numerical examples to motivate and highlight our contribution. Section III provides the details of the proposed delay aware packet scheduling algorithm. Section IV sheds light on the delay and fairness aware traffic scheduling paradigm for mixed traffic. Section V provides a discussion on the algorithmic complexities of the proposals. Section VI discusses the simulation model. Section VII showcases the results and discussion followed by the conclusion.

## II. Motivation

In this section, we would like to motivate our problem formulation and provide the rationale behind choosing the objective function to balance between fairness and QoS enhancement with an example. First, we discuss the issues related to channel aware scheduling that provide us with the motivation to include buffer status within the same framework. Thereafter, we introduce real-time traffic scheduling and demonstrate that even channel plus buffer aware scheduling is not enough when traffic with delay deadlines are considered. We provide an example with three UEs that are competing for a single RC to highlight the motivation for our proposed method. An RC is comprised of a set of six contiguous PRBs. In this example, we assume that the channel quality indicator (CQI) value does not change through the length of our example of five TTIs. We also assume that the number of bytes arriving per TTI to each of the users is deterministic and has a fixed value of 50. These assumptions are for the sake of maintaining simplicity. Later, we prove through simulations that the conclusions drawn over here are rather generic. Table I provides the relationship between a certain CQI index value and the number of bytes of data that can be transmitted with the supported type of modulation [26][27]. Table II, on the other hand, illustrates the initial conditions.

As mentioned before, only channel aware scheduling, while neglecting the buffer conditions is not optimal as it induces unfairness. The UEs closer to the eNodeB are scheduled more frequently as their channel gains are generally better. However, the UEs with poorer channel gains do not get enough transmission opportunities, which results in higher delay and packet loss [14]. In our example, only UE2, since it has the best CQI, will be scheduled in all the TTIs if such an approach is taken.

On the other hand, if channel plus buffer aware scheduling is used (Table III), the scheduling parameter becomes the minimum of the number of bytes that the channel condition allows to transmit for a particular UE and the number of bytes present in that UE's buffer[2]. This clearly increases fairness as the buffer occupancy also becomes a deciding factor. As a result, different users are selected in different TTIs (marked in bold). Hence, the overall system MAC throughput improves.

We present a few examples in the following tables to support our argument. The bold fonts in Table III-V indicate the scheduled users. The values in TTI(i+1) column of Table III-V indicate the updated values from TTI(i) column after accounting for the arriving and transmitted bytes. For example, in Table III, TTI(2) column value for UE1 is calculated by subtracting the number of bytes transmitted during TTI(1) (400) from the buffer content at the beginning of TTI(1) (400) followed by an addition of the number of newly arriving bytes (50). Similarly, since, UE2 was not selected for transmission in TTI(1), its buffer content increased by 50 in TTI(2).

In Table IV and Table V, we deal with traffic having delay deadlines[3]. In this example, we assume that 20 bytes cross delay deadline in every TTI. Channel plus buffer aware scheduling is not very effective in this regard as illustrated in Table IV. The real-time packets have delay limits after which the packets are dropped. This may result in heavy packet drop due to delay violation for the disadvantaged user. Thus, the buffer content of these "needy" users may remain low and the scheduler may not allocate them bandwidth for data transmission. As a result, the system throughput degrades over the long term and packet loss increases.

TABLE I
MAPPING OF CQI TO DATA TRANSMISSION WITHIN A TTI [26][27]

| CQI Index | Modulation | Data Transmission per PRB (bytes) | Data Transmission per RC (bytes) (1 RC=6PRBs) |
|---|---|---|---|
| 1-6 | QPSK | 42 | 252 |
| 7-9 | 16-QAM | 84 | 504 |
| 10-15 | 64-QAM | 126 | 756 |

TABLE II
EXAMPLE CQI AND BUFFER

| UE | CQI (C) | Data transmission possible based on Channel (bytes) (P) | Arrival rate (bytes/TTI) | Initial buffer size (bytes) (B) | Actual Data transmission possible (W)(bytes) $min(p_i, b_i)$ |
|---|---|---|---|---|---|
| 1 | 7 | 504 | 50 | 400 | 400 |
| 2 | 12 | 756 | 50 | 300 | 300 |
| 3 | 6 | 252 | 50 | 260 | 252 |

TABLE III
CHANNEL PLUS BUFFER AWARE SCHEDULING

| UE | Actual Data transmission possible /Buffer (bytes) | | | | |
|---|---|---|---|---|---|
| | TTI(1) | TTI(2) | TTI(3) | TTI(4) | TTI(5) |
| 1 | **400/400** | 50/50 | 100/100 | 150/150 | **200/200** |
| 2 | 300/300 | **350/350** | 50/50 | 100/100 | 150/150 |
| 3 | 252/260 | 252/310 | **252/360** | **158/158** | 50/50 |

[2] In Table III, the terms separated by "/" indicate the actual number of bytes that can be transmitted based on the buffer and channel conditions; and the buffer content respectively.

[3] In Table IV and Table V, three terms are separated by "/". They indicate the actual number of bytes that can be transmitted based on the buffer and channel conditions, the buffer content and the number of bytes to be dropped if they are not transmitted in the current TTI respectively.

TABLE IV
CHANNEL PLUS BUFFER AWARE SCHEDULING FOR REAL TIME TRAFFIC

| UE | Actual Data transmission possible/Buffer/Packets to be dropped (bytes) | | | | | |
|---|---|---|---|---|---|---|
| | TTI(1) | TTI(2) | TTI(3) | TTI(4) | TTI(5) | |
| 1 | **400/400/50** | 50/50/0 | **100/100/20** | 50/50/0 | 100/100/20 | |
| 2 | 300/300/100 | **250/250/20** | 50/50/0 | 100/100/20 | **130/130/20** | |
| 3 | 252/260/255 | 55/55/5 | 100/100/20 | **130/130/20** | 50/50/0 | |
| Total Trans-mitted | 400 | 250 | 100 | 130 | 130 | 1010 |
| Total Drop | 355 | 5 | 20 | 20 | 20 | 420 |

TABLE V
CHANNEL PLUS BUFFER AWARE SCHEDULING FOR REAL TIME TRAFFIC CONSIDERING PACKET DROP

| UE | Actual Data transmission possible/Buffer/Packets to be dropped (bytes) | | | | | |
|---|---|---|---|---|---|---|
| | TTI(1) | TTI(2) | TTI(3) | TTI(4) | TTI(5) | |
| 1 | 400/400/50 | **400/400/20** | 50/50/0 | 100/100/20 | **130/130/20** | |
| 2 | 300/300/100 | 250/250/20 | **280/280/20** | 50/50/0 | 100/100/20 | |
| 3 | **252/260/255** | 55/55/5 | 100/100/20 | **130/130/20** | 50/50/0 | |
| Total Trans-mitted | 252 | 400 | 280 | 130 | 130 | 1192 |
| Total Drop | 153 | 25 | 20 | 20 | 20 | 238 |

In our example, we can clearly see that in the first TTI, UE3 should be scheduled as it faces the danger of heavy packet drop. However, if only channel plus buffer aware scheduling is used, UE1 is selected (see Table IV). On the other hand, if we schedule UE3 based on our new objective, which subtracts the number of bytes dropped from the bytes that are actually sent (see Table V), then heavy packet drop can be avoided at the expense of lower MAC throughput. For example, we observe that if UE1 is selected, then the objective value becomes $(400 - 100 - 255 = 45)$. Here, 400 is the number of bytes transmitted by UE1 and 100 and 255 are the number of bytes dropped by UE2 and UE3 (due to high delay) respectively. Similarly, scheduling UE2 yields $(300 - 50 - 255 = -5)$ and scheduling UE3 results in $(252 - 50 - 100 = 102)$. Hence, the proposed scheduler schedules UE3. This procedure continues in every TTI. From the example, we can conclude that our proposal improves the MAC throughput while reducing the overall packet drop.

## III. DELAY AWARE REAL-TIME TRAFFIC SCHEDULING

In this section, we explain our previously proposed Dynamic Hungarian Algorithm with modification (DHAM) [14]. Thereafter, we extend DHAM to consider delay aware real time traffic and to support QoS when only a single type of traffic is being transmitted. We list the symbols used in the following sections in Table VI.

### A. MAC Throughput Maximization

CA based scheduling is an elegant solution to scheduling problems in LTE uplink. However, one must remember that CA scheduling assumes infinitely backlogged model. Unfortunately, this assumption does not always hold in real world networks. Hence, along with CA, a buffer aware scheduling mechanism must be introduced to enhance scheduling efficiency. We introduced such an algorithm (DHAM) in [14].

As explained in Section II, for DHAM, one needs to prepare the channel gain matrix ($C$) and buffer status ($B$) as shown in Table II. Please note that in Table II, we have considered a single RC, so the $C$ matrix has only one column. Multiple RC consideration will result in multiple columns. The same logic holds for $P$ and $W$ matrices. Thereafter, $P$ matrix is prepared, which signifies the number of bytes that can be transmitted (refer Table II) by mapping the elements of matrix $C$ to values found by following the MCS given in Table I. In our example shown in Table I, we have ignored the effect of coding and the corresponding $P$ matrix values are represented in bytes.

TABLE VI
SYMBOLS

| Symbols | Description |
|---|---|
| $N_U$ | Number of active UEs. |
| $M$ | Number of RCs. |
| $\Omega$ | Set of delay violating users. |
| $N_D$ | $\lvert\Omega\rvert$. |

Next, the elements of traffic matrix $W$ are prepared as follows (see Table II):

$$w_{i,j} = min(p_{i,j}, b_i) \tag{1}$$

where, $w_{i,j}$ is the element of traffic matrix $W$ that corresponds to the $i^{th}$ UE and the $j^{th}$ RC; $p_{i,j}$ is the element of matrix $P$ that corresponds to the $i^{th}$ UE and the $j^{th}$ RC (maximum possible number of bytes that the $i^{th}$ UE can send on the $j^{th}$ RC) and $b_i$ is the element of the vector B that corresponds to the $i^{th}$ UE. In (1), the $min(p_{i,j}, b_i)$ is considered because if $p_{i,j}$ is greater than $b_i$, the UE will not be able to send more data than $b_i$; since, after sending $b_i$, the UE will have no more data to send in the current TTI. On the contrary, if $p_{i,j}$ is smaller than $b_i$, the UE will be able to send at most $p_{i,j}$ due to physical channel capacity constraints.

Thus, applying equation (1) over the example of Table II, we get the final Traffic Matrix ($W$).

Utilizing the prepared $W$ matrix, the scheduling operation is done with the help of Integer Linear Program (ILP) (2-5).

$$Maximize \quad \sum_{i,j} \alpha_{i,j} w_{i,j} \tag{2}$$

$$\text{Subject to,} \quad \sum_{i,j} \alpha_{i,j} \leq M \tag{3}$$

$$\sum_{i} \alpha_{i,j} \leq 1, \forall j \tag{4}$$

$$\sum_{j} \alpha_{i,j} \leq 1, \forall i \tag{5}$$

where, $w_{i,j} \epsilon W$ is obtained from (1); $\alpha_{i,j}$ is a binary variable, which is equal to 1 if UE $i$ is allocated to RC $j$. Otherwise, $\alpha_{i,j} = 0$. The constraint (3) specifies that the number of allocations should be less than or equal to the number of RCs present. The constraint (4) implies that an RC can be allocated to only a single UE. The constraint (5) signifies that one UE can get at most one RC.

In [14], we have shown that the scheduling problem falls under the category of the well-known assignment problem and Hungarian algorithm can be applied to solve it in polynomial time. However, assignment problems require equal cardinalities of both the partites. Whereas, for the above-mentioned scheduling, $N_U$ is generally greater than $M$. Therefore, one needs to use Hungarian algorithm after adding dummy RCs to transform the optimization problem to an assignment problem. The costs/rewards of connecting UEs to dummy RCs are set to zero. Therefore, the UEs that get allocated to dummy RCs are considered to be the UEs with no allocation in the current TTI.

### B. *MAC Throughput Maximization with QoS Support*

CA-Hungarian [7] and DHAM maximizes PHY throughput and MAC throughput respectively. DHAM works optimally when used with best-effort traffic. However, when we have traffic with delay constraints, i.e., real-time traffic, DHAM may inadvertently degrade the QoS. Examples of real-time traffic are voice and video. Since, DHAM is both CA and buffer aware, it is inclined towards scheduling the UEs with better channel quality. The users with poor channel quality may not get the opportunity for transmission even when their buffer builts up. Moreover, for real-time traffic, if a packet is not transmitted within its deadline, the packet is dropped. This results in further deterioration of the QoS. Hence, DHAM requires further modifications to ensure that it works efficiently for real-time traffic. The resulting modified protocol is termed as delay aware real-time scheduling (DARTS).

DARTS uses the same traffic matrix *(W)* used for DHAM. The criterion for scheduling a user using DARTS can be summarized using the following ILP.

$$Maximize \quad \sum_{i,j} \alpha_{i,j} w_{i,j} \tag{6}$$

$$\text{Subject to,} \quad \sum_{i} \alpha_{i,j} \leq 1, \forall j \tag{7}$$

$$\sum_{j} \alpha_{i,j} \leq 1, \forall i \tag{8}$$

$$\delta_i \left(1 - \sum_j \alpha_{ij}\right) \leq D_{th} - t, \forall i \tag{9}$$

In the ILP (6-9), the constraints have inequalities in order to accept unequal number of RCs and UEs. The constraint (7) means that an RC can be allocated to at most one UE. Constraint (8) implies that an UE, if scheduled, can have at most one RC. The constraint (9) is used to ensure that a real-time packet is scheduled within its delay constraints. Here, $\delta_i$ indicates the delay of the HoL packet of the $i^{th}$ UE. If the UE has been scheduled in the current TTI then $\left(1 - \sum_j \alpha_{i,j}\right) = 0$, otherwise it is equal to 1. $D_{th}$ represents the delay deadline for the considered traffic and $t$ is the length of the TTI. Therefore, if the HoL packet of a UE is going to cross $D_{th}$ in the next TTI then $(\delta_i \leq D_{th} - t)$ is no longer satisfied. However, this condition becomes false even if a single UE has a HoL packet that is about to cross $D_{th}$ in the next TTI (whether the UE is scheduled or not). In order to avoid this situation, we add the term $\left(1 - \sum_j \alpha_{i,j}\right)$, which ensures that only the UEs that are not scheduled in the current TTI should not have a HoL packet that will cross $D_{th}$ in the next TTI. Hence, the inequality enforces scheduling of an UE when its HoL packet will cross its delay deadline if not scheduled in the current TTI.

Unfortunately, the formulation (6-9) has two serious shortcomings, which we overcome in the following sub-section.

a) *Possible infeasibility* – If $N_D > M$ then there is no means to satisfy both constraint (7) and constraint (9) at the same time. Hence, the optimization does not return any result and becomes infeasible.

b) *Inefficiency* – Constraint (9) forces a delay violating user to be scheduled. However, if the delay violating user has very poor channel conditions, scheduling that user is not advisable from the system's point of view.

### C. *DARTS Realization*

#### 1) *Case 1 – Removing possible infeasibility: When $N_U > N_D > M$*

ILP (6-9) works well as long as $N_D < M$. Otherwise, the ILP (6-9) becomes infeasible as the optimization can no longer satisfy both constraint (7) and constraint (9).

A possible solution is to formulate two ILPs. When $N_D < M$, use ILP (6-9); otherwise shortlist only the set $\Omega$ (refer Table VI) and execute DHAM on them, which has no delay constraint.

However, the problem of executing DHAM only on set $\Omega$ does not capture the number of packets that will be dropped if $\omega \epsilon \Omega$ is not scheduled in the current TTI. This is further illustrated in Table VII with a suitable example. Here, UE 2 has relatively better channel conditions but with fewer packets facing the risk of dropping. UE 1, on the other hand, has relatively poorer channel conditions but has a large number of packets about to be dropped in its buffer. Therefore, in order to enforce fairness, we should ideally prefer the scheduling of UE 1 over UE 2. However, DHAM will do the opposite. Moreover, we also need to minimize packet drop arising from the resulting scheduling. In another example that is shown in Table VIII, we can clearly see that scheduling UE1 (more critical user) leads to a drop of 298 bytes by UE1 and 500 bytes by UE2, which leads to a total drop of 798 bytes. Similarly, scheduling UE2 leads to a drop of 550 bytes. This will no doubt lead to slight unfairness, but it is better from the systems' perspective. In other words, by scheduling UE2, we perform DOM by avoiding future packet drop; thereby, controlling long term system throughput indirectly. DARTS ensures fairness over time and at the same time it enhances delay awareness without causing infeasibility.

Moreover, it is evident that if $N_D > M$, then all the $\omega \epsilon \Omega$ will not be scheduled in the current TTI. This will inevitably lead to packet drop for the users that are not scheduled. Further, even if a user is scheduled, packet drop $\left(d_{ij}\right)$ may occur due to physical channel restrictions. Therefore, in order to ensure that these users get higher priority in the $(t+1)^{th}$ TTI, the number of bytes that are dropped at the end of the $t^{th}$ TTI by UE $i$ will be added to the obtained "packets to be dropped" metric for the $(t+1)^{th}$ TTI. For long term fairness enforcement, this packet drop history can be merged with the current packet drop information using a weighted average and the result can be used in the $(t+1)^{th}$ TTI. The procedure is given in (10).

$$k_{t,i} = \vartheta \varepsilon_{t,i} + (1 - \vartheta) k_{t-1,i} \tag{10}$$

where, $k_{t,i}$ is the metric obtained for UE $i$ for $t^{th}$ TTI, $\varepsilon_{t,i}$ is the packet drop information for the $t^{th}$ TTI for UE $i$ and $\vartheta$ is a weighing factor used and has a value of 0.9. Please note, that during calculation of $k_i$ in the $(t+1)^{th}$ TTI, $\varepsilon_{t,i}$ is set to exactly the number of bytes that were dropped by the UE $i$ in the $t^{th}$ TTI. Further, from here on, we will use $k_i$ to indicate $k_{t,i}$.

TABLE VII
EXAMPLE TRAFFIC: CASE 1

| UE | Data transmission possible (bytes) | Bytes to be dropped |
|---|---|---|
| 1 | 252 | 260 |
| 2 | 504 | 5 |

TABLE VIII
EXAMPLE TRAFFIC: CASE 2

| UE | Data transmission possible (bytes) | Bytes to be dropped |
|---|---|---|
| 1 | 252 | 550 |
| 2 | 504 | 500 |

Thus, in order to ensure fairness and enforce DOM, we need to modify the objective function of the optimization procedure used in DHAM. So, the new objective function becomes –

$$\begin{aligned}
&Maximize \quad \textstyle\sum_i \sum_j \alpha_{ij}\left(w_{ij} - d_{ij}\right) - \sum_i \beta_i k_i && (11)\\
&\text{Subject to,} \quad \textstyle\sum_i \alpha_{ij} = 1, \forall j && (12)\\
&\textstyle\sum_j \alpha_{ij} \leq 1, \forall i && (13)\\
&\beta_i + \textstyle\sum_j \alpha_{ij} = 1, \forall i && (14)\\
&\alpha_{ij} = 0 \text{ } or \text{ } 1 && (15)\\
&\beta_i = 0 \text{ } or \text{ } 1 && (16)
\end{aligned}$$

where, $d_{ij}$ is the number of bytes that will be dropped by UE $i$ if UE $i$ is allocated to RC $j$. Constraint (12) signifies that an RC can be allocated to one UE while Constraint (13) denotes that a user may or may not be allocated to an RC. Constraint (14) ensures that UE $i$ is either allocated an RC ($\sum_j \alpha_{ij} = 1$) or the UE is mapped to the dummy RC ($\beta_i = 1; \sum_j \alpha_{ij} = 0$).

We replace $\left(w_{ij} - d_{ij}\right)$ by $\gamma_{ij}$ for compact representation. The problem described by Table VII has been taken care of by the $\sum_i \beta_i k_i$ term. On the other hand, consideration of $\sum_i \sum_j \alpha_{ij} d_{ij}$ handles the problem described by the example in Table VIII. The UE $i$ can report the value of $k_i$ along with the buffer status using a long buffer status report. It is impractical for the eNodeB to evaluate $k_i$ as it will require the delay values of all the packets.

Hence, ILP (11-16) effectively replaces the delay constraint of ILP (6-9) by the packet drop metric. Further, ILP (11-16) offers more user fairness than ILP (6-9) as ILP (11-16) considers all the packets that are facing the danger of dropping. However, if the UEs have very good channel conditions and danger of packet drops is absent, ILP (11-16) reverts back to ILP (2-5).

Looking closely at the objective function (11) and the constraint (14), we can identify that we can equate $\alpha_{i(M+1)} = \beta_i$ and treat $\gamma_{i(M+1)} = -k_i$. This modification makes the "packets about to be dropped if not scheduled" equivalent to a dummy RC and the dummy RC can have an inflow of $(N_U - M)$. These $(N_U - M)$ inflows are connected to the UEs that are not scheduled. For this discussion, we assume that $(N_U > M)$. Interestingly, ILP (11-16) becomes equivalent to the famous transportation problem (ILP (17-21)) with the divergence of the last dummy RC node being equal to $(N_U - M)$. So, the new formulation is given by ILP (17-21).

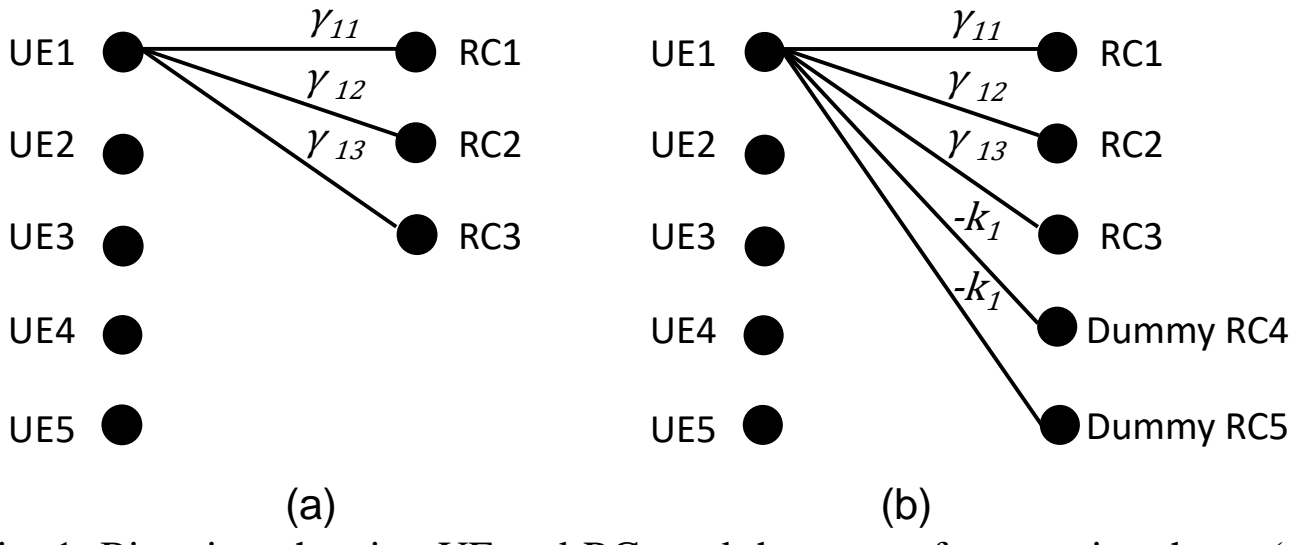

Fig. 1. Bipartites showing UE and RCs and the costs of connecting them. (a) without adding dummy RCs. (b) adding dummy RCs with same cost for each UE (assignment problem).

$$\begin{aligned}
&Maximize \quad \textstyle\sum_i \sum_{j=1}^{M+1} \alpha_{ij}\gamma_{ij} && (17)\\
&\text{Subject to,} \quad \textstyle\sum_i \alpha_{ij} = 1, \forall j = 1,2,\dots,M && (18)\\
&\textstyle\sum_i \alpha_{ij} = N_U - M, \forall j = M+1 && (19)\\
&\textstyle\sum_j \alpha_{ij} = 1, \forall i && (20)\\
&\alpha_{ij} = 0 \text{ } or \text{ } 1 && (21)
\end{aligned}$$

Constraint (19) forces the inflow of the dummy node to be $(N_U - M)$. The other constraints are same as ILP (2-5).

Finally, in order to convert ILP (17-21) to an equivalent assignment problem (ILP (22-25)), we have to replicate the dummy RC node to create $(N_U - M)$ dummy RCs and restrict the inflow of each of the dummy RCs to 1, which modifies constraint (19) to $\left(\sum_i \alpha_{ij} = 1, \forall j = M+1, M+2, \dots, N_U\right)$. In graphical terms, we will create as many dummy nodes as required to make the cardinalities of both the partites equal. The cost of connecting to each of the dummy nodes will be same for a certain UE. Fig. 1 is a graphical representation of the creation of the dummy RCs. We have shown the edges for only one UE for brevity. The reader should visualize that every node of the left partite is connected to every element of the right partite in similar manner. Thus, ILP (11-16) is converted to an equivalent ILP (22-25). The entire process preserves optimality. This formulation allows us to use the low complexity Hungarian algorithm to solve our problem.

$$\begin{aligned}
&Maximize \quad \textstyle\sum_i \sum_{j=1}^{N_U} \alpha_{ij}\gamma_{ij} && (22)\\
&\text{Subject to,} \quad \textstyle\sum_i \alpha_{ij} = 1, \forall j = 1,2,\dots,M,(M+1),\dots,N_U && (23)\\
&\textstyle\sum_j \alpha_{ij} = 1, \forall i && (24)\\
&\alpha_{ij} = 0 \text{ } or \text{ } 1 && (25)
\end{aligned}$$

where, constraint (23) captures the conditions of constraints (18) and (19) of ILP (17-21). We consider $N_U > M$ where $1,2,\dots,M$ are the actual RCs and $(M+1),\dots,N_U$ denote the dummy RCs.

*2) Case 2 – Applying DARTS when $N_U > M > N_D$*

The scheduling of UEs if $\mathrm{N_U} > \mathrm{M} > \mathrm{N_D}$ uses the same ILP (22-25).

#### 3) *Case 3 – Applying DARTS when $M > N_U \geq N_D$*

For the previous discussions, we have considered that $M < N_U$. However, another case may arise where $M > N_U$. In such a scenario, if we allow only one RC per UE, some RCs will be wasted which could have been used by the active UEs. Allocating multiple RCs to a single UE is in line with the carrier aggregation ideology of LTE-A. Please note that incorporating carrier aggregation results in a sub-optimal solution.

In order to achieve carrier aggregation, we use ILP (11-16) in an iterative manner for this purpose. However, we omit the $d_{ij}$ term for this purpose. The term $d_{ij}$ is obtained through preprocessing by considering the fact that a certain UE will be allocated to a single RC. But, in this scenario, there is a scope for allocation of multiple RCs to a single UE. Hence, the calculation of $d_{ij}$ becomes dependent on the allocations arising from multiple iterations. As a result, including $d_{ij}$ will lead to suboptimal allocations. We prove through simulations that the effect of using $d_{ij}$ is minimal in such a scenario.

Moreover, $k_i$ is made equal to the number of bytes that will be dropped if UE $i$ is not scheduled in the current TTI $\left(k_i = \varepsilon_{t,i}\right)$. We can clearly see that as long as the number of UEs is less than the number of RCs considered for allocation, the term $\sum_i \beta_i k_i$ does not play any role in the allocation. At the end of every iteration, the values of $k_i$ are updated by subtracting the number of bytes that will be transmitted by UE $i$ due to the allocation obtained in the current iteration. The minimum value that $k_i$ can attain is equal to zero. In the final iteration, the updated value of the term $\sum_i \beta_i k_i$ ensures that the delay violating UEs may be exclusively considered. The value of $w_{i,j}$ is also updated in accordance with the allocation. It should be noted that since $N_U < M$, dummy UEs with zero buffer content are added in this case for framing an assignment problem. We present the steps for DARTS implementation in Algorithm 1.

ALGORITHM 1 DARTS AND DAFS

***Input*** : $p_{ij}, b_i, k_i, \varepsilon_{t,i}, M, N_D, N_U, k_{t-1,i}$
***Output*** : *Scheduling decision variable* $\alpha_{ij}$

1. $w_{i,j} \leftarrow min\left(p_{i,j}, b_i\right)$;
2. $d_{ij} \leftarrow max\left(\varepsilon_{t,i} - w_{i,j}, 0\right)$;
3. *Calculate* $k_{t,i}$ *using equation* (10);
4. *if* $(N_U > N_D > M)$ *or* $(N_U > M > N_D)$
5.     *Execute ILP*(22 – 25);
6. *end if*;
7. *if* $N_U < M$
8.     $d_{i,j} \leftarrow 0$;
9.     $k_i \leftarrow \varepsilon_{t,i}$;
10.     *while* (*Unscheduled RCs available*)
11.         *Execute ILP*(11 – 16);
12.         $k_i \leftarrow max\left(k_i - \sum_j \alpha_{ij} w_{i,j}, 0\right)$;
13.         $b_i \leftarrow max\left(b_i - \sum_j \alpha_{ij} w_{i,j}, 0\right)$;
14.         *if RC j is not scheduled yet*
15.             $w_{i,j} \leftarrow min\left(p_{i,j}, b_i\right)$;
16.         *else*
17.             $w_{i,j} \leftarrow 0 \ \forall i$;
18.         *end if*;
19.     *end while*;
20. *end if*;

**Please note, $\varepsilon_{t,i}$ is generated using equation (26) for DAFS. All other steps are same.*

## IV. DELAY AND FAIRNESS AWARE TRAFFIC SCHEDULING FOR MIXED TRAFFIC

In this section, we propose a delay-aware fair scheduling algorithm for mixed traffic (DAFS) that maximizes the MAC throughput while meeting minimum QoS constraints when multiple traffic types are being scheduled. Thereafter, we propose priority flipping at the UE to further mitigate lower priority traffic starvation.

### A. *Delay-aware fair scheduling algorithm for mixed traffic executed at the eNodeB*

In Section III, a single class traffic with delay deadline was considered. However, one must look into multiple classes of traffic because a single user may generate voice, video and data packets at the same time. These different types of traffics have varying QoS requirements. For example, voice generally has stringent delay bounds ($< 50$ ms) for the air interface [17]. Video, on the other hand, has to satisfy relatively relaxed delay deadlines ($< 150$ ms) [17]. Finally, data has no delay constraints. However, data packets may suffer from bandwidth starvation if the real time traffic load is sufficiently high. Hence, we need to take care of each of the traffic types individually in order to provide overall user satisfaction; which demands that we must feed parameters relating to each of the traffic types in the optimization engine. However, allowing too many parameters to enter into the optimization algorithm makes the convergence time impractical from the implementation point of view. Therefore, we construct a single parameter that ensures the individual delay bounds of each class present in the considered traffic mix. This further keeps the execution time of the optimization algorithm within practical limits. Our modified metric $(\varepsilon_{t,i})$ for $i^{th}$ user for $t^{th}$ TTI is given below,

$$\varepsilon_{t,i} = m_i^{vo} + m_i^{vi} + m_i^{d} \tag{26}$$

where, $m_i^{vo}$= the number of voice packets that will be dropped, if the user is not scheduled in the current TTI; $m_i^{vi}$= the number of video packets that will be dropped if the user is not scheduled in the current TTI; and $m_i^{d}$= the number of bytes in the buffer that is above a pre-defined buffering threshold ($B_{Th}$ in bytes).

Thereafter, $k_i$ is calculated using equation (10). This $k_i$ value is used in ILP (22-25) for implementing the multi-class DAFS. This ensures that a UE with urgent requirements is always preferred for scheduling. Following the model used in DARTS, the UE will report the value of $k_i$ along with the buffer status to the eNodeB. Please refer to Algorithm 1 for the implementation procedure of DAFS.

### B. *DAFS with priority flipping at the UE*

In this sub-section, we mitigate lower priority traffic starvation. In contrast to the algorithms discussed so far, where the processing is done in the eNodeBs and UEs employ strict priority packet forwarding, DAFS with priority flipping (DAFS-PF) employs additional optimization at the UE.

In strict priority transmissions, the order of priority followed is voice, video and then data. Unfortunately, in strict priority transmission, if the higher priority traffic composition is sufficiently high then there is constant arrival of traffic in the higher priority buffers. This results in deferring of the lower priority packet transmission as the lower priority packets are transmitted only when all the higher priority traffic buffers are empty. Hence, the lower priority packets may suffer from starvation.

Therefore, in DAFS-PF, the UE may increase the priority value of the lower priority packets if it senses the development of a possible starvation. In order to achieve this, we associate rewards to packets and our target is to maximize the rewards while filling up the allocated bandwidth. This requirement matches with the requirements of a knapsack problem. The problem is formally defined as follows.

$$Maximize \sum_i \sum_j \alpha_{ij} r_{ij} \quad (27)$$

$$\text{Subject to,} \quad \sum_i \sum_j \alpha_{ij} w_{ij} \leq G \quad (28)$$

$$\alpha_{ij} = 0 \; or \; 1 \quad (29)$$

where, $j \in \{voice, video, data\}$, $\alpha_{i,j}$ is a binary variable indicating the scheduling information of the $i^{th}$ packet of the $j^{th}$ type, $r_{i,j}$ is the reward associated with the $i^{th}$ packet of the $j^{th}$ type, $w_{i,j}$ is the size of the $i^{th}$ packet of the $j^{th}$ type and $G$ is the allocated bandwidth. Next, we explain the design of the reward variables that suits our purpose.

$$r_{i,voice} = \frac{D_{i,vo}}{D_{Th,vo}}, r_{i,video} = \frac{D_{i,vi}}{D_{Th,vi}}, \quad (30)$$

$$r_{i,data} = \begin{cases} \left(\frac{B_D}{B_C}\right)\left(\frac{B_c - B_{Th}}{B - B_{Th}}\right) & if \; B_c > B_{Th} \\ 0 & otherwise \end{cases} \quad (31)$$

where, $D_{i,vo}$ is the $i^{th}$ voice packet delay, $D_{i,vi}$ is the $i^{th}$ video packet delay, $D_{Th,vo}$ is the voice delay threshold, $D_{Th,vi}$ is the video delay threshold, $B_D$ is the data buffer occupancy, $B_c$ is the current buffer occupancy, $B_{Th}$ is the buffer threshold and $B$ is the buffer capacity.

The explanations of the reward metrics are as follows: For both voice and video, as the delay of a packet approaches the delay threshold, the reward for transmitting the packet increases. Therefore, if video is getting starved due to high input of voice packets, at some point of time $r_{i,video}$ will be greater than $r_{i,voice}$ and hence the video packet will get transmitted prior to the voice packets.

However, data packets are not of the class of real time traffic and therefore, cannot be associated with delay. Hence, we choose buffer build up as the metric for prioritizing data packets. Since, data packets can be delayed, they are not given any priority when $B_c < B_{Th}$. However, when $B_c > B_{Th}$ and data packets are present in the buffer, we can say that the data packets are being starved due to the presence of other higher priority packets. Therefore, the data packets are "*marked*" and same reward is associated to each and every marked data packet. The priority among the data packets follows first-in-first-out principle. Therefore, as the buffer fills up, the priority of the data packets increases, and they become more likely to be transmitted. Thus, DAFS-PF seeks to mitigate lower priority traffic starvation.

In order to minimize the execution time of the algorithm, these metric calculations must be a part of preprocessing before every scheduling instance. During the scheduling instance, only the knapsack algorithm is executed. The design of the algorithm will be such that all the metrics (rewards) will be recalculated while packet transmissions in the preceding TTI are ongoing. This helps in minimizing scheduling delay.

## V. A Note on Algorithmic Complexity

The creation of the traffic matrix has a worst case complexity of $O(mn)$, where $m$ is the number of RCs and $n$ is the number of UEs [28]. Therefore, for creating a symmetric matrix, the complexity becomes $O(max(m,n)^2)$.

The traffic creation matrix of DARTS and DAFS is $O\big(m(n+1)\big)$. The Hungarian algorithm assignment complexity is $O(max(m,n)^3)$ [29]. Thus, the complexity is $O(max(m,n)^3) + O(max(m,n+1)^2)$.

The user side action of DAFS-PF has to perform a user procedure following the knapsack optimization paradigm. The optimization solution has been obtained through the greedy method as there is packet fragmentation provision in LTE. The resulting complexity is $O(n \log n)$ [28].

## VI. Simulation Model

The simulation studies have been carried out on a system level simulator developed in the OMNeT++ network simulator. We have considered a seven-cell scenario (Fig. 2), where a single cell is surrounded by six first tier cells. The studies have been carried out on the center cell, which acts as the serving cell and the first-tier cells provides the interfering signal power. All the results obtained have been recorded in the center cell.

The UEs have been deployed in the cells by following a Poisson point process. In this paper, we have assumed that all the UEs are directly connected to its serving eNodeB. Uplink transmission power control and MCS have been considered as per the guidelines provided in [27]. We have used a block fading channel model, where the channel conditions remain constant over a TTI [15][33]. The work assumes that the UEs are not mobile. However, the work is perfectly valid without any modifications for users with mobility; the mobility being restricted by LTE specifications. If mobile scenario is considered, the control message exchange will increase as the speed of the UE increases. Further, we assume that the eNodeB has knowledge of the CQIs, buffer lengths and critical packets of all the UEs at the time of taking the scheduling decisions. For changing the input voice load, we have varied the generation interval of the packets of a two-state Markov VoIP Model [30]. Similarly, for the video model [31], we have altered the frames per second in order to vary the load. The load of the self-similar data source has been varied following the method given in [34]. The simulation parameters are listed in Table IX and Table X. The results have been obtained by solving the developed optimization problems using the Hungarian algorithm in OMNET++.

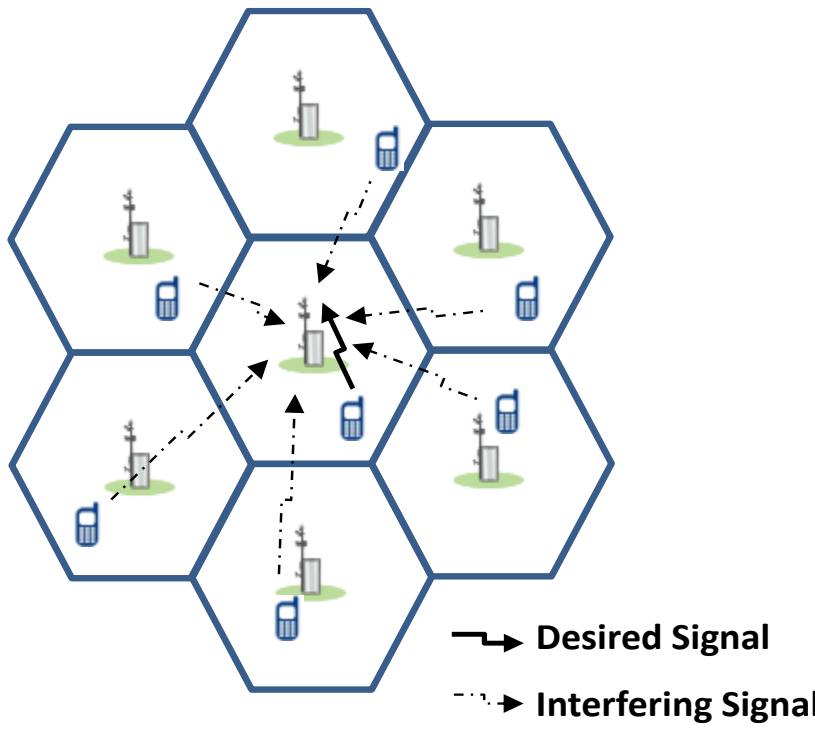

Fig. 2. Simulation Model

TABLE IX
TRAFFIC MODELS

| Parameter | Values |
|---|---|
| Voice traffic model | two-state Markov [30] |
| Voice packet size | 40 bytes |
| Silence indicator (SID) packet size | 15 bytes |
| Video traffic | Near real time video [31] |
| Frame per second | 15 |
| No. of packets in a frame | 8 |
| Min. video frame size | 1.5 kilo bytes |
| Packet size | Truncated Pareto; K=40 bytes, α=1.2, mean=50 bytes, max=250 bytes |
| Packet inter-arrival time | Truncated Pareto; K=2.5 ms, α=1.2, mean= 6 ms, max= 12.5 ms |
| Data traffic | Self-similar [32] |
| Packet payload | uniformly distributed between [46,1500] bytes |

TABLE X
LTE PARAMETERS

| Parameter | Value |
|---|---|
| Scenario | UMa |
| Inter-site distance | 500 m |
| System bandwidth | 10 MHz |
| Center frequency | 2 GHz |
| No. of prbs ($n_{PRB}$) | 50 (48 for data) |
| No. of prbs in a RC | 6 |
| Path loss ($PL$) model | Non line of sight |
| Shadowing standard deviation | 4 dB |
| UE max transmit power ($P_{max}$) | 24 dBm |
| Uplink power control (PC) | $max\ (P_{max}, P_o + \alpha PL 10 \log_{10} n_{PRB})$ |
| $\alpha$ for PC | 1.0 |
| $P_o$ For PC | -106 dBm |
| UE distribution | Poisson point process (PPP) |
| Simulation duration | 10 seconds (10,000 TTIs) |
| MCS | QPSK, 16-QAM, 64-QAM with varying code-rates as given in [30] |

## VII. RESULTS AND DISCUSSION

This section illustrates the performance of the proposed algorithms, namely DARTS, DAFS and DAFS-PF. In our previous work, we have extensively studied DHAM [14]. In that work, we have compared the performance of DHAM with that of recursive maximum expansion [13], channel aware Hungarian algorithm [7] and buffer based channel dependent scheduler [15]. In [14], we have established that DHAM provides better MAC throughput, lower buffer build-up and more fairness as compared to the other scheduling protocols. Therefore, in this paper, we compare our newly proposed algorithm for single class traffic (DARTS) with DHAM only. However, for multi-class traffic, we have compared our proposed algorithms (DAFS and DAFS-PF) with DHAM and APASS.

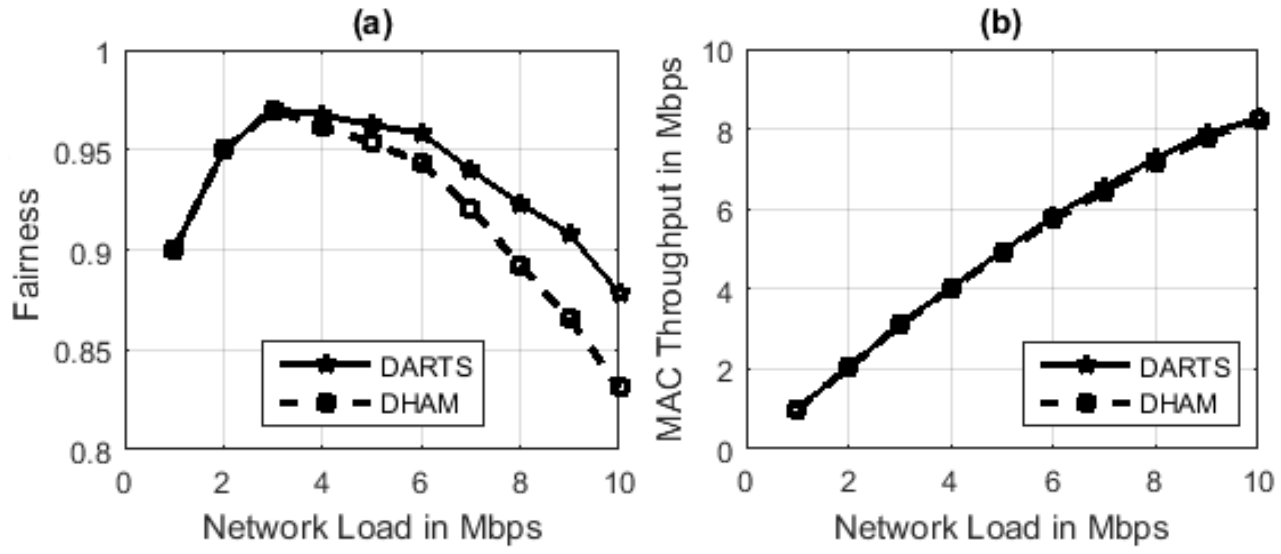

Fig. 3. (a) Fairness vs Network Load (b) MAC Throughput vs Network Load

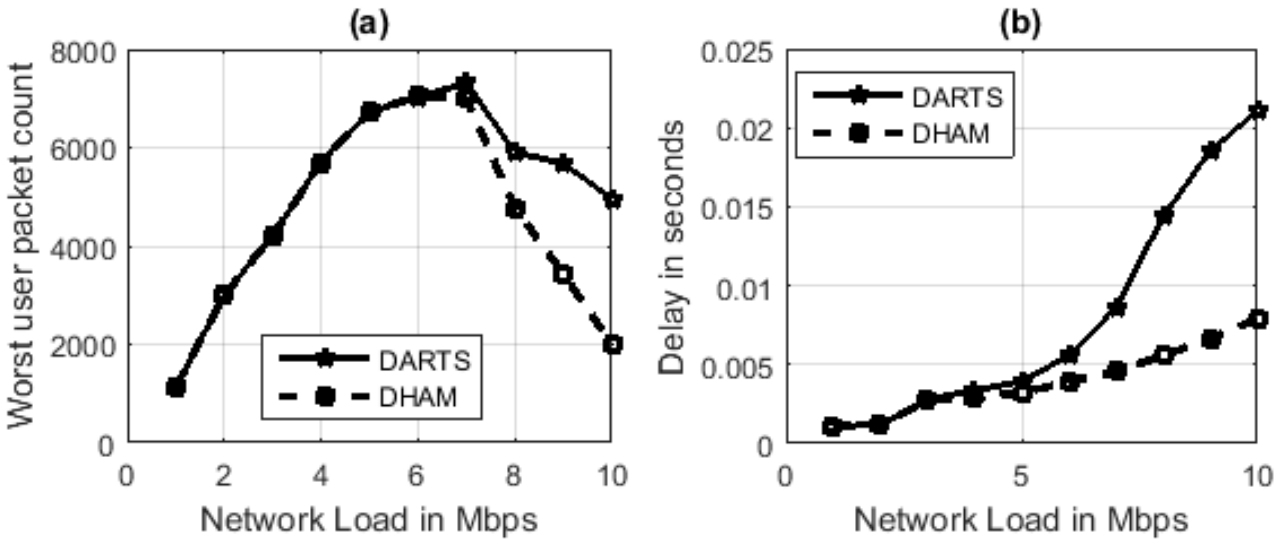

Fig. 4. (a) Number of worst user voice packets delivered vs Network Load (b) Voice delay vs Network Load

### A. *Delay aware real-time scheduling*

In this sub-section, we compare the performance of DARTS with DHAM. The main parameters to study are the fairness, the MAC throughput, the number of packets delivered by the user having worst channel conditions and the delay.

DARTS performs much better in terms of fairness as depicted in Fig. 3a. Fairness is improved in case of DARTS because whenever a user suffers from bad channel conditions for a considerable amount of time, DARTS provides opportunity for them to occupy the channel. Whereas, DHAM only seeks to improve the MAC throughput and as a result generally selects the users with better channel quality. However, it should be noted that the improvement on fairness is obtained without any sacrifice on MAC throughput as seen in Fig. 3b. This happens as DARTS reduces packet drop due to delay. Also note that the fairness improves initially before degrading again. We can observe such a trend because the fairness depicted here is based on the opportunity that the users get for occupying the channel. Hence, when the load is very low, the utility function defined by (1) remains low for poor channel users both because of poor CQI and low buffer occupancy and they get the opportunity to transmit only when the users with good channel conditions have almost empty buffers. Once scheduled, the poor channel users empty their buffers and must wait again till their buffer gets filled to a certain extent so as to possess a better utility function than the good channel users. However, the buffers get filled up at a slower rate in low load. As load increases, the buffer fill-up rate also increases and thus fairness improves. However, after a certain point, the good channel users again start to get privileges because their good CQI as the buffers simply fill up faster and hence, their probability of getting scheduled is increased.

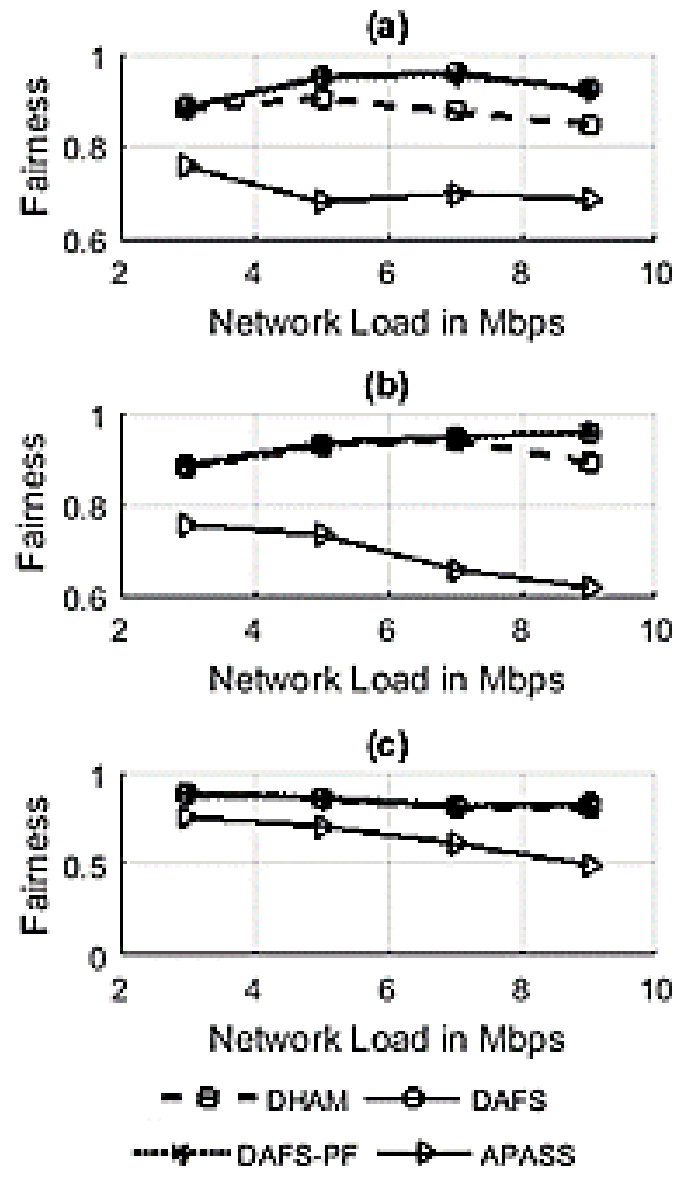

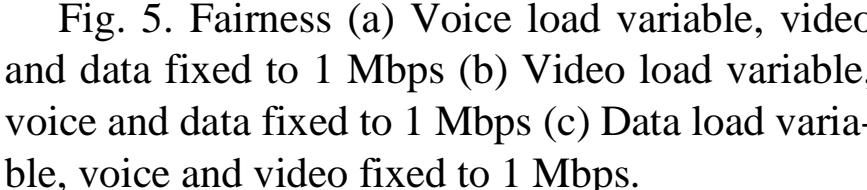

Fig. 5. Fairness (a) Voice load variable, video and data fixed to 1 Mbps (b) Video load variable, voice and data fixed to 1 Mbps (c) Data load variable, voice and video fixed to 1 Mbps.

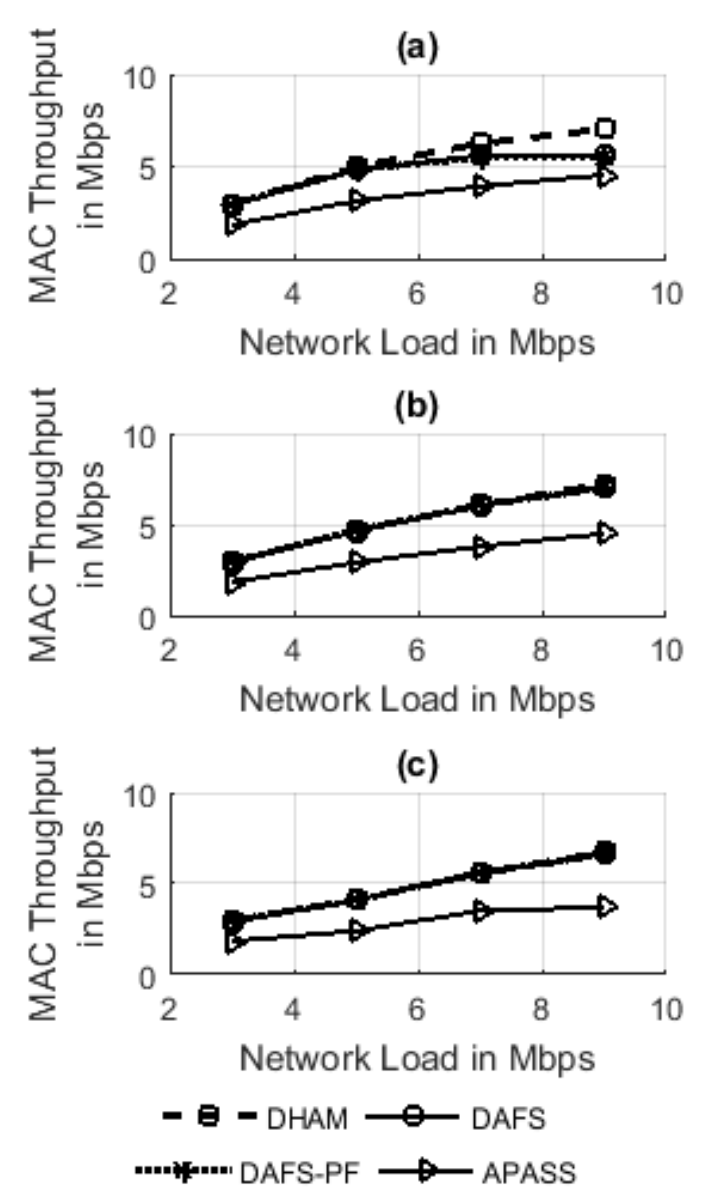

Fig. 6. MAC throughput (a) Voice load variable, video and data fixed to 1 Mbps (b) Video load variable, voice and data fixed to 1 Mbps (c) Data load variable, voice and video fixed to 1 Mbps

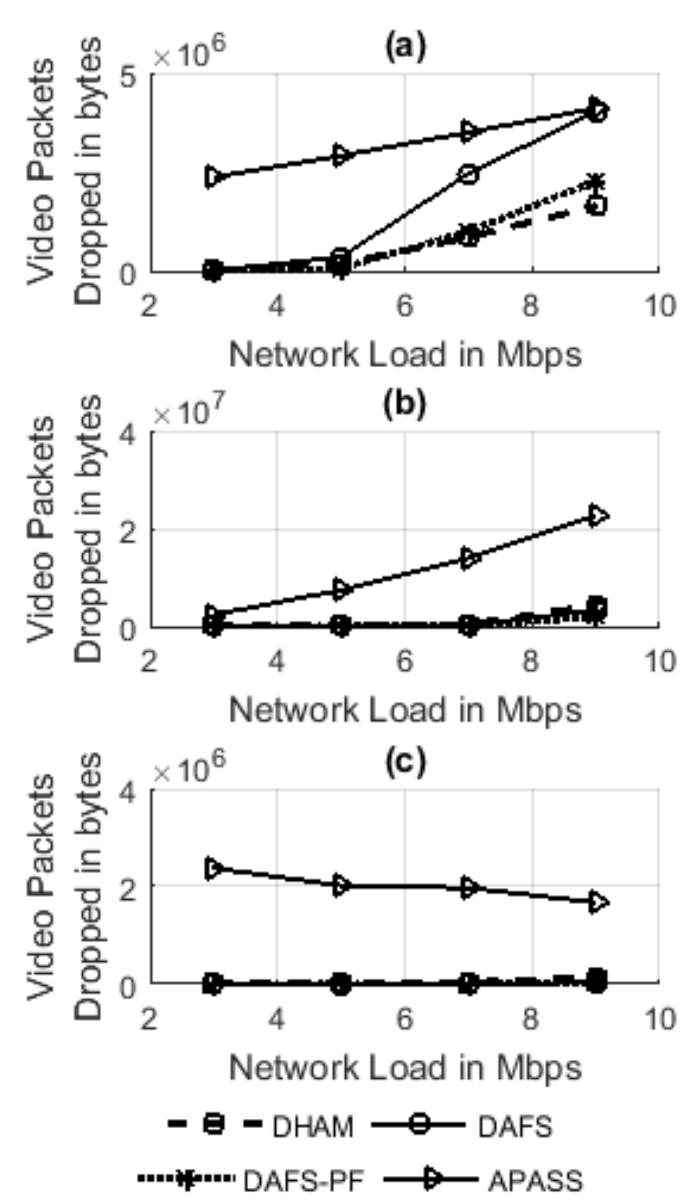

Fig. 7. Video packets dropped (a) Voice load variable, video and data fixed to 1 Mbps (b) Video load variable, voice and data fixed to 1 Mbps (c) Data load variable, voice and video fixed to 1 Mbps

The scheduling principle of DARTS prioritizes users who have packets about to violate the delay constraints. Therefore, if the input rates of all the users are similar, the poor channel quality users are normally worst hit. Fig. 4a, illustrates the enhancement in the delivery of packets for a user having the worst channel condition due to the delay aware scheduling of DARTS. In Fig. 4a, we can see that the packet delivery of the worst user improves initially. However, as the load increase beyond 70%, the higher requirement of the good channel users makes it beneficial for the system to schedule them as it improves the overall system throughput. Hence, the packet delivery count of the worst channel user degrades after a certain load.

Finally, in order to prefer the users with higher number of delayed packets, DARTS defers the transmission from the users with better channel quality and therefore increases the average delay as illustrated in Fig. 4b. However, DARTS ensures that the delay remains within the delay limits.

### B. *QoS enhancement of multiple classes of traffic*

This sub-section displays the effects of the DAFS and DAFS-PF when multiple classes of traffic are present in the traffic mix. DAFS is the multiclass version of DARTS and therefore, we have not included the results of DARTS in this sub-section. DAFS and DAFS-PF has been compared against the performance of DHAM and APASS. Note that, for the result compilation, we have considered the DAFS version with packet drop history. For each of the following figures, we have three sub-plots [a, b and c]. For sub-plot (a), the voice load is made variable while the video and data load are fixed at 1Mbps. For the sub-plot (b), the video load is altered while the voice and data load are fixed at 1Mbps. Finally, for the last sub-plot (c), voice and video load are fixed to 1 Mbps, while data load is varied.

As seen in Fig. 5, the fairness provided by DAFS and DAFS-PF is higher than that of both DHAM and APASS. This happens because, both DAFS and DAFS-PF give more priority to the users that have poorer channel conditions. APASS, on the other hand, removes schedulable users in a heuristic manner in its second stage. As a result, the optimal fairness is not attained. However, enhancement of fairness results in degradation of MAC throughput as can be seen from Fig. 6. MAC throughput is higher for DHAM as it seeks to maximize MAC throughput. However, DAFS and DAFS-PF seeks to maximize MAC throughput while meeting delay constraints. As a result, the MAC throughput delivered by them is comparatively lower. On the other hand, the semi-heuristic nature of APASS makes it sub-optimal even in terms of MAC throughput.

Fig. 7 reveals that the DAFS leads to higher video packet drop as compared to DHAM when the percentage of voice traffic is higher in the traffic composition (see Fig. 7a). This directly follows from the fact that DAFS yields lower MAC throughput. Moreover, strict priority scheduling clears out the voice buffer before transmitting the video packets. As a result, video buffer gets filled up quickly when the voice traffic load is high. DAFS-PF tries to rectify this shortcoming of DAFS by the application of priority flipping. However, DAFS-PF also yields lower MAC throughput than DHAM and hence drops higher number of video packets as compared to DHAM. However, the performance of APASS is poorer than all of our proposals as it ignores several bandwidth starving users because of the usage of heuristic utility functions. On the other hand, when the percentage of voice is low in the traffic mix, DHAM, DAFS and DAFS-PF gives comparable performance (see Fig. 7b and 7c).

However, when it comes to the number of video packets transmitted by the worst channel user, DAFS and DAFS-PF clearly outperforms DHAM. We observe that in this regard, DAFS-PF performs the best because of priority flipping. The comparison can be seen in Fig. 8.

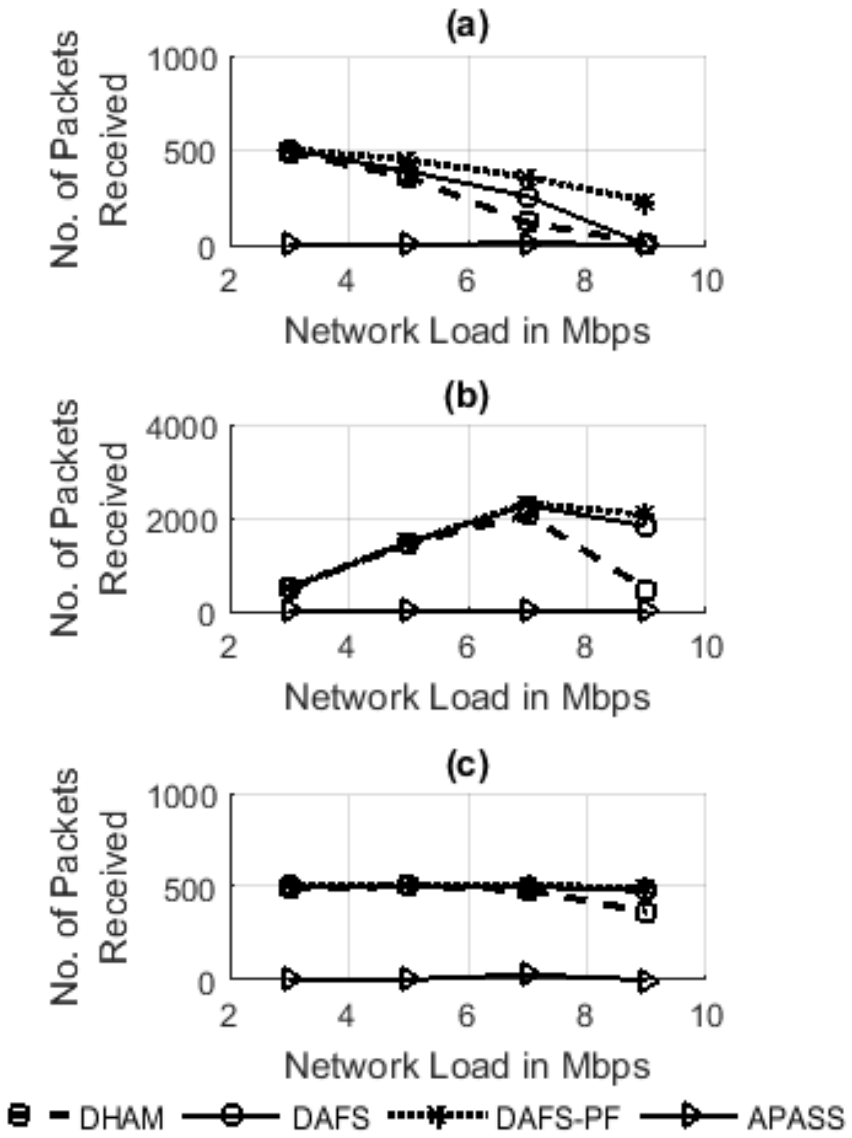

Fig. 8. Number of worst user video packets delivered (a) Voice load variable, video and data fixed to 1 Mbps (b) Video load variable, voice and data fixed to 1 Mbps (c) Data load variable, voice and video fixed to 1 Mbps.

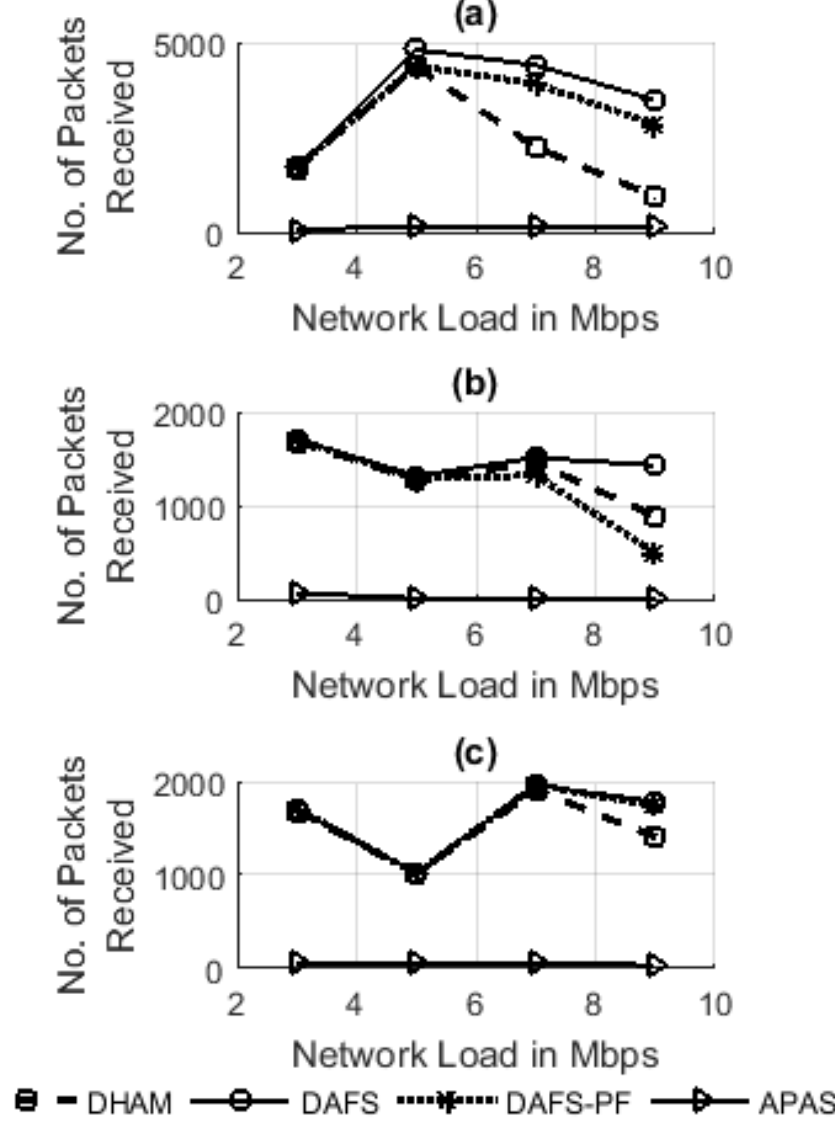

Fig. 9. Number of worst user voice packets delivered (a) Voice load variable, video and data fixed to 1 Mbps (b) Video load variable, voice and data fixed to 1 Mbps (c) Data load variable, voice and video fixed to 1 Mbps

The number of video packets delivered by the worst user decreases as load increases because the lack of transmission opportunities results in higher packet drop. This happens as all the UEs accumulate large number of packets in their queues when the load is high. However, the video performance of the DAFS-PF is improved by borrowing the bandwidth that was to be used for transmitting voice packets in DAFS. Hence, the worst user voice performance is poorer for DAFS-PF than it is for DAFS as shown in Fig. 9. Nevertheless, both the algorithms perform better than DHAM for heavy load conditions because of their fair scheduling policy. APASS, however, shows significant poor scheduling for the worst channel user. The primary reason for such an observation is the heuristic utility function used in APASS, which takes a reactive approach for tackling delay outage. On the other hand, we proactively minimize the delay outage. Moreover, we intentionally delay the transmission of packets from users with good channel quality while staying within the delay constraints so as to allocate the vacated bandwidth to the poor channel users. As a result, our proposals perform better than APASS. Additionally, we believe that APASS is inefficient in a static scenario as the mean path-loss of the users do not change in such a situation. However, our optimal choice of the utility function makes it efficient in a static scenario.

Finally, an interesting trend can be observed in Fig. 9b and Fig. 9c; the number of voice packets delivered when video or data load is varied while keeping the voice load fixed falls till the network load of 5Mbps (Fig. 9b and Fig. 9c). At this moderate load, the worst user does not get enough selection preference as the voice packets are dropped due to delay overshoot. However, as the load increases further, the fairness is invoked due to higher accumulation of video and data packets. Finally, as the load increases even further, all the users become heavily loaded and compete for the available bandwidth. Therefore, the voice packet reception from the worst user decreases again.

## Conclusion

In this paper, we have demonstrated that channel aware scheduling alone in the LTE/LTE-A uplink should not be the de-facto mechanism for resource allocation. In order to utilize the physical resources efficiently, one must make a cross layer optimization including information from the MAC layer. The use of buffer state information while scheduling is obligatory to increase the effective throughput, i.e., the MAC throughput.

However, this crude consideration may be rendered unfair when real-time traffic is used. Therefore, we have proposed DARTS for a single class of real time traffic, DAFS and DAFS-PF for multiple classes of real-time traffic. These advanced low complexity scheduling algorithms produce optimal results while effectively enhancing the fairness among users. The novelty of the algorithms lies in the usage of delay outage minimization in place of HoL delay. Further, these algorithms make use of the long buffer status reports to transmit the packet drop value from the UEs to the eNodeB. This relieves the system from transmission of heavy control information. Finally, priority flipping, introduced in DAFS-PF, boosts video performance and reduces data starvation in appropriate scenarios.

The RC assignment part of the algorithms has been implemented using the Hungarian algorithm and extensive simulation studies have been performed using the formulated Integer Linear Programs. The studies have confirmed that the uplink performance can be improved significantly by incorporating the proposed schedulers in the eNodeB.

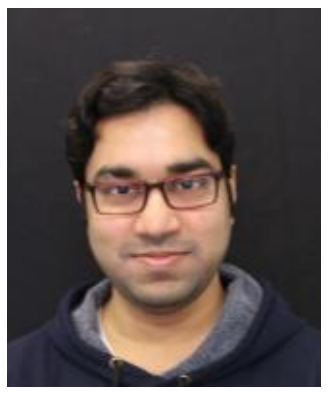

**Atri Mukhopadhyay** received his Ph.D. degree from the Indian Institute of Technology, Kharagpur, India in 2019. He is presently working as a post-doctoral researcher in Trinity College Dublin, Ireland. His research interests include wireless-optical integrated networks, optical access networks, wireless access networks and edge computing.

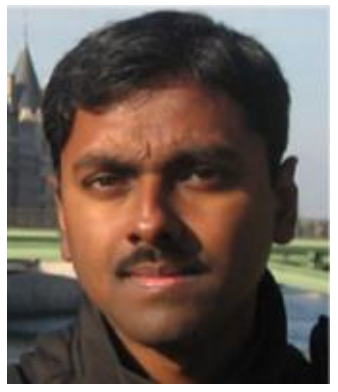

**Goutam Das** has obtained his Ph.D. degree from the University of Melbourne, Australia in 2008. He has worked as a postdoctoral fellow at Ghent University, Belgium, from 2009-2011. Currently he is working as an Assistant Professor in the Indian Institute of Technology, Kharagpur. His research interest is in the area of optical access networks, optical data center networks, radio over fiber technology, optical packet switched networks and media access protocol design for application specific requirements.